\documentstyle[11pt,a4]{article}
\begin{document}
\newcommand{\be}{\begin{equation}}
\newcommand{\ee}{\end{equation}}
\newcommand{\ben}{\begin{eqnarray}}
\newcommand{\een}{\end{\eqnarray}}
\newcommand{\nn}{\nonumber}
\newcommand{\no}{\noindent}
\newcommand{\n}{\label}
\def \p {\partial}

\title{Varying c and particle horizons}

\author{
Luis P Chimento\dag\, Alejandro S Jakubi\dag\, and Diego Pav\'{o}n\ddag\
\\
\\
\dag\ {\small Departamento de F\'{\i}sica, Universidad de 
Buenos Aires, 1428~Buenos Aires, Argentina}\\
\ddag\ {\small Departamento de F\'{\i}sica, Universidad Aut\'onoma de 
Barcelona, 08193 Bellaterra, Spain
\footnote{Fax: + 34-93 5812155; Electronic address: diego@ulises.uab.es}}\\
}

\maketitle

\date{}

\begin{abstract}

We explore what restrictions may impose the second law of thermodynamics
on varying speed of light theories. We find that the attractor scenario 
solving the flatness problem is consistent with the generalized second 
law at late time.

\end{abstract}

\[\]
\[\]

\newpage
\section{Introduction}

Recently proposals were advanced to solve the horizon and flatness problems 
of the standard big--bang cosmology -in a different way that inflationary
picture does- as well as the cosmological constant problem by allowing the
speed of light in vacuum and the Newtonian gravitational constant to vary with
time \cite{moffat,albrecht,barrow,cg}. These approaches are collectivelly 
called varying speed of light (VSL) theories.
Possible variations of the fundamental physical constants in the expanding
Universe are currently of particular interest because of the implications of
unified theories, such as string theory and M-theory
\cite{marciano84,barrow87,damour94}. They predict that additional compact
dimensions of space exist. The ``constants" seen in our three--dimensional
subspace of the theory will vary according to any variation in the scale
lengths of the extra compact dimensions. While other scenarios with
varying fundamental constants have been considered, like scalar--tensor
theories of gravity (see, e.g., \cite{nordtvedt}) prescribing that $G$ must be
function of a scalar field (the Brans--Dicke field) and the varying fine
structure constant theory of Bekenstein \cite{bekenstein}, they do not touch
the speed of light and respect Lorentz invariance. By contrast,  certain
fundamental theories, including strings, could admit spontaneous violation of
CPT and Lorentz invariance \cite{kostelecky}. For instance, within string 
theory, quantum aspects of the interactions between particles and 
non--perturbative quantum fluctuations break supersymmetry and Lorentz 
invariance \cite{ellis}. It became an interesting issue to investigate the 
violation of Lorentz invariance in  high energy phenomena \cite{coleman}. 
VSL theories also break Lorentz invariance rendering their approach 
non--covariant. They provide simple effective models to describe 
these effects.

In this letter we investigate what constraints the second law of
thermodynamics may bring on the formulation of VSL theories, a point
that to the best of our knowledge have received no attention so far.
As it turns out these constraints are strong for homogeneous and 
isotropic spacetimes lacking of a particle horizon. However, for 
spacetimes possessing a particle horizon the restrictions are much 
less severe -at least at late time.

\section{Field equations and constant attractor solution}

Let us consider a expanding Friedmann--Lema\^{\i}tre--Robertson--Walker 
(FLRW) universe whose source of the gravitational field is a perfect fluid 
and assume that the speed of light in vacuum is not really a constant 
but varies time in some unknown manner, i.e., $c = c(t)$. The 
corresponding generalized ``Einstein field equations" for a 
homogeneous and isotropic universe can be written as 

\be
H^{2} = \frac{8\pi G\rho }3-\frac{k c^2(t)\ }{a^2},
\n{fr} 
\ee

\be
\frac{\ddot{a}}{a} = -\frac{4\pi G}3\left[\rho +\frac{3P}{c^2(t)}\right],
\n{ac}
\ee
\noindent
where $H \equiv \dot{a}/a$ is the Hubble factor, and $k \, (=+1, 0, -1)$ 
denotes the curvature of the spatial sections. Equation (\ref{fr}) implies 
that the energy density is not conserved as the universe expands  

\be
\dot \rho +3 H \left(\rho +\frac {P}{c^2}\right)=\frac{3k}{4 \pi G} \, 
\frac{c\dot c}{a^2}.
\n{cons}
\ee

It therefore looks like as though the Universe were an ``open system"
in the thermodynamical sense. Here we shall explore some of its 
consequences.

It is generally accepted that the present matter density of the 
universe is below the critical value \cite{bahcall} -though voices 
of dissent can be heard \cite{rowan}. The density parameter $\Omega$, 
defined as the ratio of the energy density of the universe with the 
critical density, $\Omega \equiv 8\pi G\, \rho/(3 H^{2})$, is one of 
the best-studied cosmological parameters and its low value is 
indicated by a number of independent methods for the 
study of clusters of galaxies. They include the
mass-to-light ratio, the baryon fraction, the cluster abundance and the 
mass power spectrum. Thus, if the energy density of our Universe were
dominated by clustered matter we would find the problem that a universe 
with $\Omega_{0}\simeq {\cal O}(1)$ requires aextreme fine tuning of 
initial conditions. This is the flatness problem, and it can find a 
solution within the VSL framework without invoking inflationary 
fields. By using the above $\Omega$ expression in (\ref{fr}),
differentiating it with respect to time and resorting to 
(\ref{cons}), the evolution equation 

\begin{equation}
\dot{\Omega} = (\Omega -1) \left[\left(1 + \frac{3P}{\rho c^{2}}\right)
+ \frac{2 \dot{c}}{c}\right] \equiv f(\Omega)
\label{domega}
\end{equation}

\noindent
follows. For $\dot{c} \neq 0$ it has two constant solutions, namely 
$\Omega = 1$, which is unstable, and $\Omega_{*}$. The latter arises when the 
square parenthesis in (\ref{domega}) vanishes. Our interest focuses on it 
because it is stable -since $\p [f(\Omega)/\p \Omega]_{\Omega_{*}} < 0 $-,
and is an attractor of the system. For $\Omega = \Omega_{*}\, $ the speed of 
light obeys the law

\begin{equation} \label{ct0}
c(a)=c_{1} \, a^{\Omega_{*}} \exp\left[-\frac{3\Omega_{*}}{2}\int
dt\,\left(1+\frac{P}{\rho c^2}\right)  H\right].
\end{equation}

\noindent 
We note that this expression leads to a decreasing speed of light
provided that the dominant energy condition holds. Equation   
(\ref{cons}) can be solved by using (\ref{ct0})

\begin{equation} \label{rhot}
\rho(t)= \frac{\rho_{1}}{G}\, a^{2\left(\Omega_{*}-1\right)}
\exp\left[-3\Omega_{*}\int
dt\,\left(1+\frac{P}{\rho c^{2}}\right) H\right],
\end{equation}

\noindent
hence it follows that $\rho=\rho_{1} \, c^2/(G c_{1}^2a^{2})$. By 
combining it with (\ref{fr}) a relationship between the density 
parameter and the integration constants can be obtained, namely

\begin{equation} \label{omega0}
\Omega_{*}= \left[1-\frac{3kc^2_{1}}{8\pi\rho_{1}}\right]^{-1},
\end{equation}

\noindent
and by virtue of (\ref{fr}) the scale factor can be written 
in terms of the speed of light

\begin{equation} \label{at0}
a(t)= \sqrt{\frac{k}{\Omega_{*}-1}}\int c(t)\,dt.
\end{equation}

In the particular case of a linear barotropic equation of state
$P = \left(\gamma-1\right)\rho c^2(t)$ with constant adiabatic index 
$\gamma$, it follows that 

\begin{equation} \label{ctb}
c(a)=c_{1}\, a^{\beta},
\end{equation}

\begin{equation} \label{Omega0b}
\Omega_{*} = \frac{2\beta}{2-3\gamma},
\end{equation}
\noindent

\noindent
where use of (\ref{ct0}) has been made.

\noindent 
For ordinary fluids the strong energy condition (SEC) holds. This
implies $\gamma>2/3$, thereby $\beta<0$ for $\Omega_{*}>0$. From 
(\ref{at0}) the scale factor is

\begin{equation} \label{atb}
a(t)=\left[\left(1-\beta\right)c_{1}\sqrt{\frac{k}{\Omega_{*}-1}}\, t\right]
^{\frac{1}{1-\beta}}\equiv a_{1} t^{\frac{1}{1-\beta}}.
\end{equation}

\section{Entropy considerations}
Let us assume that the number of particles in a comoving volume
is conserved  (i.e., $ N \equiv n a^{3} =$ constant), then the particle 
number density obeys  
\be
\dot{n} + 3 H n = 0.
\n{pnd}
\ee
This combined with Gibbs equation

\be
n T \dot{s} = \dot{\rho} - \left(\rho + \frac{P}{c^{2}}\right) 
\frac{\dot{n}}{n},
\n{gibbs}
\ee
where $s$ is the entropy per particle and $T$ the fluid temperature, 
leads to

\be
nT \dot{s} = \dot{\rho} + 3 H \left(\rho + \frac{P}{c^{2}}\right).
\n{gibbs1}
\ee
From (\ref{cons}) and (\ref{gibbs1}) it follows that

\be
n T \dot{s} = \frac{3 k}{4\pi \, G} \, \frac{c \dot{c}}{a^{2}}.
\n{gibbs2}
\ee

\noindent
Note that the entropy variation implied by last equation cannot
be attributed either to dissipative processes (since the fluid
is perfect) or particles production (for $N$ is a constant). 
We are led to conclude that the variation of $c$ entails that
the entropy of the fluid must vary. This may be justified  
-at least naively. An increase in $c$ means a widening of the 
past light cone of the observers. Automatically they acquire
more information and the entropy decreases accordingly 
\cite{information}. That is to say, 
$\dot{c} < 0  \Longrightarrow \dot{s} > 0$ as well as
$\dot{c} > 0 \Longrightarrow \dot{s} <0$. This together with 
(\ref{gibbs2}) implies that $c$ cannot increase in open universes 
(k = -1), and that flat and closed universes don't admit
$\dot{c} \neq  0$.

The consequences are rather restrictive for cosmological models 
with varying speed of light. The only admisible FLRW model with 
$\dot{c} \neq 0$ is the open one. Obviously, one may always 
introduce some traditional source of entropy such a viscous 
dissipation (only that in such a case the fluid is no longer 
perfect), or perhaps particle production from the quantum vacuum;
but this complicates matters and we wish to keep the discussion as
simple as possible.

\noindent
Note, by passing, that not every fluid is consistent with $\dot{c} \neq 0$.
Think, for instance, in a radiation fluid -equation of state
$P = \rho c^{2}/3 $. There one has \cite{pathria} 
\[
s = \frac{\rho +(P/c^{2})}{nT} = \frac{4 \rho}{3 n T} = \frac{8}{45}\, 
\frac{\pi k_{B}}{\zeta(3)},
\]  
and accordingly $\dot{s} = 0$. Therefore, in view of (\ref{gibbs2})
-barring a flat FRW- a pure radiation fluid cannot act as the gravitational 
source of a cosmology with $\dot{c} \neq 0$. \\

As is well--known, particle horizons may occur quite naturally in cosmological 
models and these have associated an entropy by a formula formally identical 
to that of event horizons (either black hole or cosmological)
\cite{gibbons} 

\be
S_{H} = \frac{k_{B}}{4}\, \frac{A}{l_{Pl}^2},
\n{sh}
\ee

\noindent 
where $k_{B}$ is the Boltzmann constant, $l_{Pl} \equiv (G\hbar/c^{3})^{1/2}$
the Planck's length, and $A$ the area of the horizon. The latter is given
by $A = 4 \pi \, l_{H}^{2}$, with 

\be
l_{H} = a(t) \int_{0}^{t} \frac{c(t')}{a(t')} \, \mbox{d}t' .
\n{lh}
\ee
Particle horizons exist provided the integral does not diverge. The
rationale behind attaching an entropy to a particle horizon is that
the area is a measure of the lack of knowledge of the observer about 
the conditions prevailing in the universe beyond the horizon.

\noindent
If a FLRW universe filled with a perfect fluid has a particle horizon,
the generalized second law (GSL) of thermodynamics (firstly devised for 
black holes in causal contact with its environment and later extended 
to cosmological settings) states that the entropy in the fluid enclosed 
by the horizon plus the entropy of the horizon cannot decrease in time 
\cite{paul}

\be
\dot{S}_{f}+ \dot{S}_{H} \geq 0.
\n{gsl1}
\ee

\noindent
Here $S_{f} = (4 \pi/3) l_{H}^{3} \, ns$. It is natural to expect that
(\ref{gsl1}) restricts the temporal dependence of $c$ less severely 
than the corresponding expression in the absence of horizons (i.e., 
when $S_{H} = 0$). Taking into account that 
$\dot{l}_{H} = H l_{H} + c $, and that $(l_{H}/a)^{.} = c/a$,
the GSL takes the form

\begin{equation}
4 \pi N c \left(\frac{l_{H}}{a}\right)^{3}\left[\frac{s}{l_{H}} +
\frac{k \dot{c}a}{4\pi NTG}\right] + \frac{\pi k_{B}}{G \hbar} \, c^{2}
\left[(2c H + 3 \dot{c}) l_{H}^{2} + 2 c^{2} l_{H}\right] \geq 0 .
\n{gsl2}
\end{equation}
 
\noindent 
To draw specific consequences of last equation, we use the constant 
attractor solution $\Omega = \Omega_{*}$ given by (\ref{ctb}) and 
(\ref{atb}). As we are considering just classical fluids and the 
horizon entropy is semiclassical in nature we may leave aside 
any consideration of an early quantum phase. Hence to restrict 
ourselves to the classical era we replace the lower index of the 
integral in (\ref{lh}) by some initial time  $t_{cl}\, (> 0) $ 
which corresponds to the commencement of the the aforesaid era. 
As a consequence 

\begin{equation} \label{lhc}
l_{H} = \frac{c_{1}}{a_{1}^{-\beta}}\, t^{1/(1-\beta)}\ln\frac{t}{t_{cl}}
\end{equation}

\noindent 
remains finite and a horizon exists. One may held the view that by
introducing a lower cutoff we illegitimately provide a horizon 
to an  otherwise horizon--free universe. In keeping with that view the very 
restrictive consequences for $c(t)$ spelled above should apply. By
contrast, the more liberal view that the cutoff is admissible since any
observer travelling backward in time will eventually hit the 
quantum era (in which -presumably- the space--time ceases to be a 
continuum and the observer should see a foam--like structure with
the light cones taking random orientations \cite{moffat1}),
gives a reasonable chance to relax those consequences.\\

To obtain $\dot {S}_{f}$ we must must know the temperature 
evolution. The latter is governed by \cite{lima}

\be\label{dT}
{\dot{T}\over T}=
-3H\left({\p p/\p T \over\p\rho/\p T}\right)_{\!n} +
{n\dot{s}\over(\p\rho/\p T)_n},
\ee

\noindent
therefore a positive specific entropy variation implies that in an expanding
universe the temperature will decrease more slowly with a declining speed of
light. Here we will consider two limiting cases at late time: monoatomic 
nonrelativistic mater, and radiation.

(i) In the first case, $\rho = mn+(3/2)nT$ and $P/c^2 = nT$, equation 
(\ref{dT}) reduces to

\begin{equation} \label{dTnr}
{\dot{T}\over T}=
-2H\left(1-\frac{k\left(\Omega_{*}-1\right)^2}{4\pi G n_{1} T}\, 
a^{2} \dot{a} \ddot{a}\right),
\end{equation}

\noindent
and its general solution is

\begin{equation} \label{Ttnr}
T(t)=T_{1}t^{-2/(1-\beta)}+T_{2} t^{2(2\beta-1)/(1-\beta)},
\end{equation}

\noindent where  $n_{1}$, $T_{1}$ are positive constants and $T_{2}$
depends on the previously defined parameters. In the large time limit the
homogeneous part becomes dominant (i.e., $T\propto a^{2}$ exactly like in a
constant speed of light cosmology), and combination with (\ref{gibbs2}) leads
to

\begin{equation} \label{stnr}
s(t)=s_{1} t^{(2\beta+3)/(1-\beta)}.
\end{equation}

\noindent 
Then, for $t\to\infty$, $S_{H} \propto t^{5/(1-\beta)}\ln^{2} t$,
$S_f\propto t^{(2\beta+3)/(1-\beta)}\ln^3 t$ and $S_H$ dominates over $S_f$,
so that (\ref{gsl1}) is satisfied.

(ii) In the second case, $\rho=c_{2} T^{4}$ and $P =\rho c^2/3$, equation
(\ref{dT}) becomes

\begin{equation} \label{dTr}
\frac{\dot {T}}{T}=- \left[1-\frac{3k\left(\Omega_{*}-1\right)^2}
{4c_{2}GT^4}\frac{\dot a^{2}\ddot {a}}{a}\right] H,
\end{equation}

\noindent
with general solution

\begin{equation} \label{Ttr}
T(t)=\left[T_1t^{4/(1-\beta)}+T_2 t^{2(2\beta-1)/(1-\beta)}\right]^{1/4},
\end{equation}

\noindent 
where  $T_{1}$ is a positive constant and $T_{2}$ depends on the
previously defined parameters. For $t\to\infty$ two cases arise. When
$-1<\beta<-1/2$, $T\propto 1/a$ as in standard cosmology, and when
$-1/2<\beta<0$, one has $T\propto a^{(2\beta-1)/2}$. In the first case
$S_f\propto t^{2(1+\beta)/(1-\beta)}\ln^3 t$, while in the second $S_f\propto
t^{(2\beta+3)/2(1-\beta)}\ln^3 t$. Again $S_{H}$ dominates $S_{f}$ and the GSL
is satisfied in both instances.

\section{Concluding remarks}
We have seen that the second law of thermodynamics implies rather severe
restrictions on $c(t)$ in FLRW cosmologies free of particle horizons. 
Specifically, in open universes $c(t)$ cannot augment, and in flat and 
closed universes $c$ must stay constant. Nonetheless, the presence of 
a particle horizon renders the situation less acute. In particular, for 
the constant atractor solution of section II, the GSL is fulfilled at 
late time both for non-relativistic monoatomic fluids and extreme 
relativistic fluids. A similar study for other cosmological 
solutions should be a worthy undertake.\\

This work has been partially supported by the Spanish Ministry of 
Science and Technology under grant BFM 2000-0351-C03-01. 
L.P.C. and A.S.J. thank the University of Buenos Aires for 
partial support under project TX-93.

\end{document}